\documentclass[reprint,showpacs,amsmath,amssymb,superscriptaddress]{revtex4-2}

\usepackage{graphicx}
\usepackage{dcolumn}
\usepackage{bm}
\usepackage{changes}

\makeatletter

\begin{document}


\title{Ion trap with gold-plated alumina: substrate and surface characterization} 



\author{Myunghun Kim}
\thanks{These authors contributed equally to this work.}
\affiliation{Department of Electrical Engineering, Pohang University of Science and Technology (POSTECH), 37673 Pohang, Korea}
\author{Keumhyun Kim}
\thanks{These authors contributed equally to this work.}
\affiliation{Department of Electrical Engineering, Pohang University of Science and Technology (POSTECH), 37673 Pohang, Korea}
\author{Jungsoo Hong}
\affiliation{Department of Electrical Engineering, Pohang University of Science and Technology (POSTECH), 37673 Pohang, Korea}
\author{Hyegoo Lee}
\affiliation{Department of Electrical Engineering, Pohang University of Science and Technology (POSTECH), 37673 Pohang, Korea}
\author{Youngil Moon}
\affiliation{Department of Electrical Engineering, Pohang University of Science and Technology (POSTECH), 37673 Pohang, Korea}
\author{Wonchan Lee}
\affiliation{Department of Physics, Pohang University of Science and Technology (POSTECH), 37673 Pohang, Korea}
\author{Sehyun Kim}
\affiliation{Pohang Accelerator Laboratory (PAL), Pohang University of Science and Technology (POSTECH), 37673 Pohang, Korea}
\author{Taekyun Ha}
\affiliation{Pohang Accelerator Laboratory (PAL), Pohang University of Science and Technology (POSTECH), 37673 Pohang, Korea}
\author{Jae-Yoon Sim}
\affiliation{Department of Electrical Engineering, Pohang University of Science and Technology (POSTECH), 37673 Pohang, Korea}
\author{Moonjoo Lee}
 \email{moonjoo.lee@postech.ac.kr}
\affiliation{Department of Electrical Engineering, Pohang University of Science and Technology (POSTECH), 37673 Pohang, Korea}

\date{\today}


\begin{abstract}
We describe a complete development process of a segmented-blade linear ion trap.
Alumina substrate is characterized with an X-ray diffraction and loss-tangent measurement.
The blade is laser-micromachined and polished, followed by the sputtering and gold electroplating.
Surface roughness is examined at each step of the fabrication via both electron and optical microscopies.
On the gold-plated facet, we obtain a height deviation of tens of nanometers in the vicinity of the ion position.
Trapping of laser-cooled $^{174}$Yb$^{+}$ ions is demonstrated.
\end{abstract}


\maketitle 


\section{Introduction}

Trapped atomic ion is a preeminent system for developing a quantum computer~\cite{Leibfried03, Monroe13}. 
Nowadays, quantum gate operation and qubit measurement can be performed at state-of-the-art levels~\cite{Ballance2016, Gaebler2016, Crain2019}.
Several quantum algorithms and error-correction codes were implemented recently~\cite{Debnath2016, Stricker2020, Erhard2021}, which would be extended with more ion qubits in the near future.
As well, in order to realize quantum advantage~\cite{Arute2019, Wu2021} or to solve more practical problems~\cite{Montanaro2016, Moll2018} with the ions, it is needed to develop a trap device in which a very large number of qubits can be controlled and measured at high fidelities~\cite{Kielpinski02, Monroe2014}.
To this end, great efforts have been devoted with various engineering technologies~\cite{Siverns2017a, Romaszko2020}.
For example, microelectromechanical system (MEMS) techniques were employed to develop miniaturized ion traps~\cite{Wright2013, Hong2016, Lee2019a, Auchter2022a}, cryogenics offered a stable environment to the ions~\cite{Poitzsch1996, Labaziewicz08, Pagano2018}, and complementary metal-oxide semiconductor (CMOS) technology enabled high-resolution, repeatable production of the ion-trap hardware~\cite{Mehta2014, Stuart2019, Blain2021}. 

The scalable ion traps would be broadly categorized into two groups~\cite{Niedermayr2014}: according to the substrate, semiconductor-based materials or other dielectrics.
In case of the former, on a substrate like Si/SiO$_2$~\cite{Britton2009,  Wilpers2012, Sterling2014, Niedermayr2014} or GaAs/AlGaAs~\cite{Stick2006}, a series of well-established microfabrications, such as etching, lithography, and metal deposition, were done to develop a trapping chip. 
This approach enabled the generation of ``High Optical Access'' trap~\cite{Maunz2016}, and would also be compatible with the CMOS and very-large-scale integration (VLSI) technologies. 
In other dielectric substrates, such as alumina~\cite{Hensinger2006, Blakestad2009, Kienzler15, Kaufmann2017, Lu2019}, fused silica and quartz~\cite{Seidelin06}, and diamond~\cite{Brewer2019, Teller2021}, remarkable achievements were reported including a junction trap~\cite{Hensinger2006, Blakestad2009}, transport gates~\cite{Clercq2016, Kaufmann2017}, and stable trap operation at cryogenic temperatures~\cite{Pagano2018}.
Differently from the former, the fabrication on such dielectrics/ceramics would not precisely follow those of standard semiconductor engineering, therefore certain development steps, like laser machining~\cite{Ragg2019}, gold electroplating~\cite{Blakestad2010, Kaufmann2017a}, and its surface control~\cite{Hou2021}, have been actively investigated in the ion-trap research: It is needed to provide detailed, complete information about the non-standard semiconductor engineering---this consists of the first motivation of our work.

The second motivation is associated with the fundamental properties of the substrate, affecting the ions' characteristics. 
For instance, the dielectric loss, a bulk property of the substrate, causes power dissipation of the driving radio frequency (RF) voltages. 
This would result in blackbody radiation (BBR) shift of a qubit transition~\cite{Safronova2011, Nordmann2020}, which is very detrimental for, e.g.~ion-clock experiments~\cite{Huntemann2016, Brewer2019}.
In parallel, surface quality of the substrate  is also crucial to the ions~\cite{Hite2012}. 
Flat facets make the electric field uniform, and reduce the surface adsorbate and impurities that are potential sources of electric-field noise~\cite{Lin2016a, Hou2021}.
Hence, it is important to characterize both the bulk and surface properties of the substrate for precise ion-trap quantum experiments.
Here, we characterize both the bulk and surface properties of our substrate, present a detailed fabrication process of a segmented-blade ion trap, and trap laser-cooled $^{174}$Yb$^{+}$ ions.
X-ray diffraction (XRD) spectroscopy shows that our substrate is a crystalline $\alpha$-alumina, and the loss tangent is measured over a broad frequency range.
The alumina is machined with a laser to make the holes and grooves at the micrometer scale, succeeded by the sputtering of titanium and gold layers.     
Electroplating of gold follows to obtain a gold layer with a thickness of about 3~$\mu$m. 
After each fabrication step, we assess the surface roughness with a scanning electron microscope (SEM) and high-resolution optical microscope. 
The roughness is further quantified through a confocal optical microscopy, from which we obtain a height deviation of gold surface as small as $< 30$~nm, at a location close to ions. 
We proceed to assemble an ion-trapping system with the blades, and load laser-cooled $^{174}$Yb$^{+}$ ions successfully. 
An experimental plan is outlined to associate the substrate properties with the characteristics of ions.

The novelty of our work is pointed out in the following ways. 
First, in the ion-trap research, it is for the first time that the bulk properties of an actual substrate are directly characterized with the XRD and loss-tangent measurements, as far as we are aware. 
Second, it is also an original approach that surface quality is quantified at every step of the fabrication, which gives sequential information for the final shape of a device.
Finally, we hereby provide a full, detailed description of developing an ion-trap system that exploits a dielectric substrate.

\begin{figure} [!t]
	\includegraphics[width=3.3in]{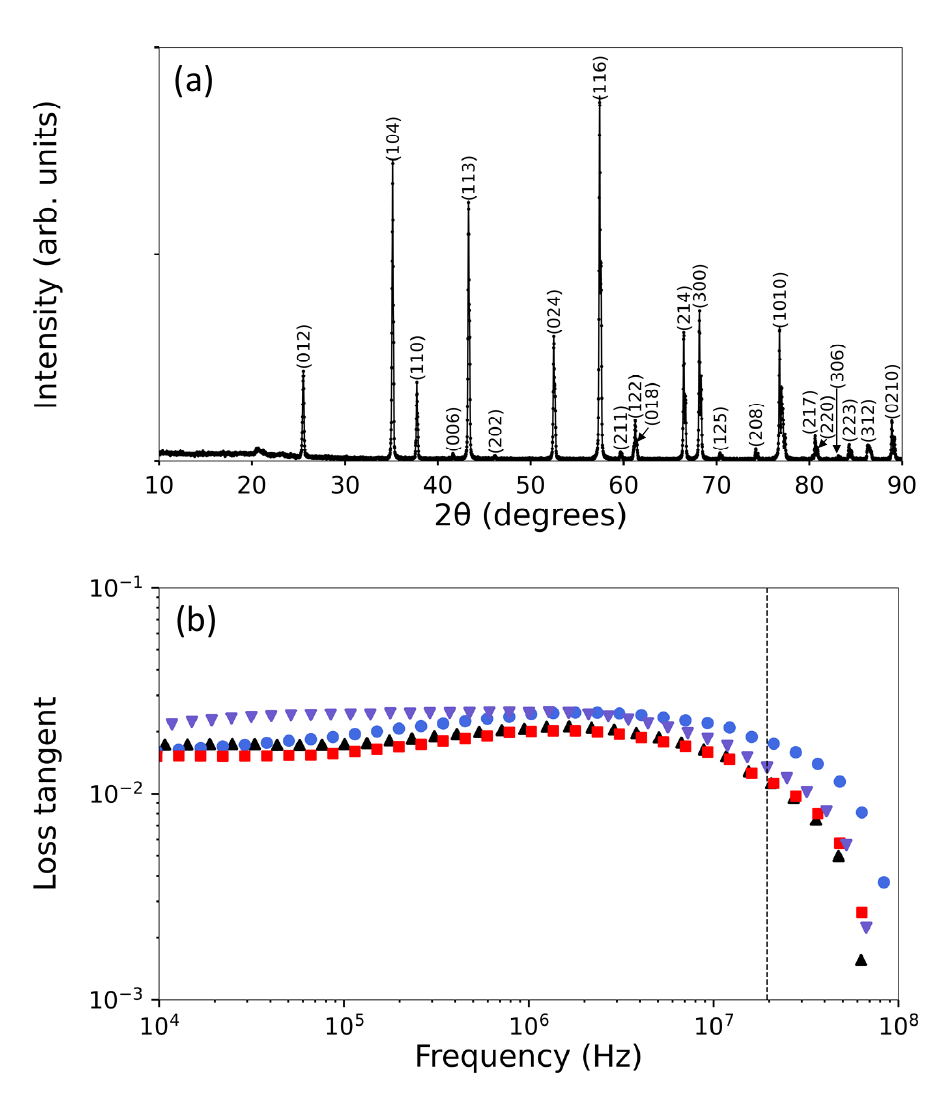} 
	\caption{	
	(a) X-ray diffraction spectrum of alumina. 
	Diffracted intensity as a function of twice the sample rotation angle $\theta$.
	(b) Measured loss tangent of alumina. 
	Four samples (denoted by different colors and symbols) are measured at frequency range from 10~kHz to 100~MHz. 
        The vertical dashed line indicates our trap operating frequency 19.525~MHz.
	The origins of errors are described in the main text.
	}
	\label{fig:xrd_loss_tangent}
\end{figure}

\begin{figure} [!t]
	\includegraphics[width=3.3in]{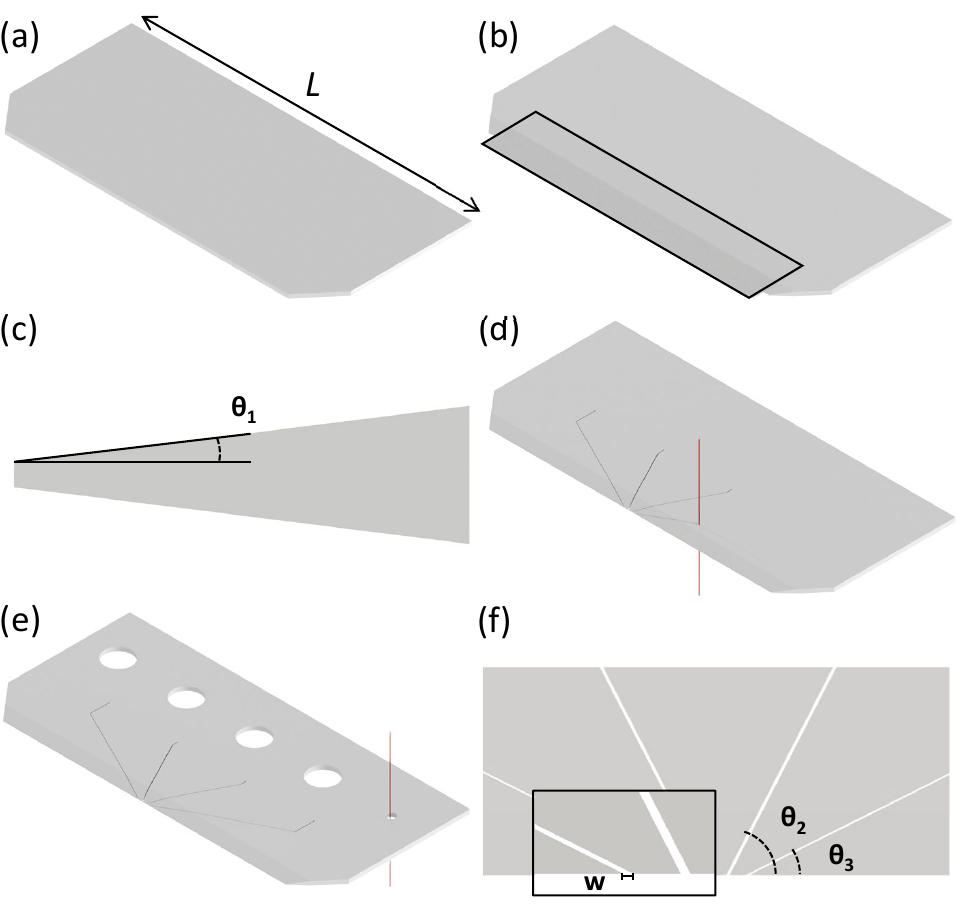} 
	\caption{
	Machining steps of alumina blade.
	(a), (b) Substrate with a thickness of 300~$\mu$m and $L=25.2$~mm.
	Blade is shaped and tapered with diamond grinding wheel.
	(c) Tapered corner. $\theta_{1}=8.0^{\circ}$.
	Four grooves in (d) and five holes in (e) are micromachined with laser. 
	(f) Central area with $\theta_{2}=62.6^{\circ}$, $\theta_{3}=26.5^{\circ}$, and $w=50\,\mu$m.	
	}
	\label{fig:fab_step}
\end{figure}

\section{Trap fabrication}    

\subsection{Alumina blade}

We consider the segmented-blade ion trap of Refs.~\cite{Hucul2015a, An2017, He2021} and note that, in our trap here, auxiliary biasing rods are installed for compensating the micromotion~\cite{Hong2022}.
The blade body is made of sintered alumina (Al$_{2}$O$_{3}$), an often used substrate in ion-trap platforms~\cite{Hensinger2006, Blakestad2009, Kienzler15, Kaufmann2017, Lu2019}. 
Alumina has been employed in the ion-trap research for several reasons: low RF losses reduce heat dissipation in the trap~\cite{Vila1998}, large bandgap of several eVs~\cite{French1990, Filatova2015} precludes the light-induced charge carrier effect~\cite{Lakhmanskiy2019}, and it is feasible to engineer the substrate through pulsed or continuous wave (CW) laser machining~\cite{Samant2008, Yan2011, Preusch2014}. 
Moreover, alumina is a ceramic with relatively high thermal conductivity ($20\sim30$~W/(m$\cdot$K)), and is chemically stable such that one can perform nano/microfabrication reliably. 
We hereby perform an XRD spectroscopy and loss-tangent measurement of a bulk sintered alumina substrate, ex situ and prior to additional processing.

\begin{figure*} [!t]
	\includegraphics[width=6.3in]{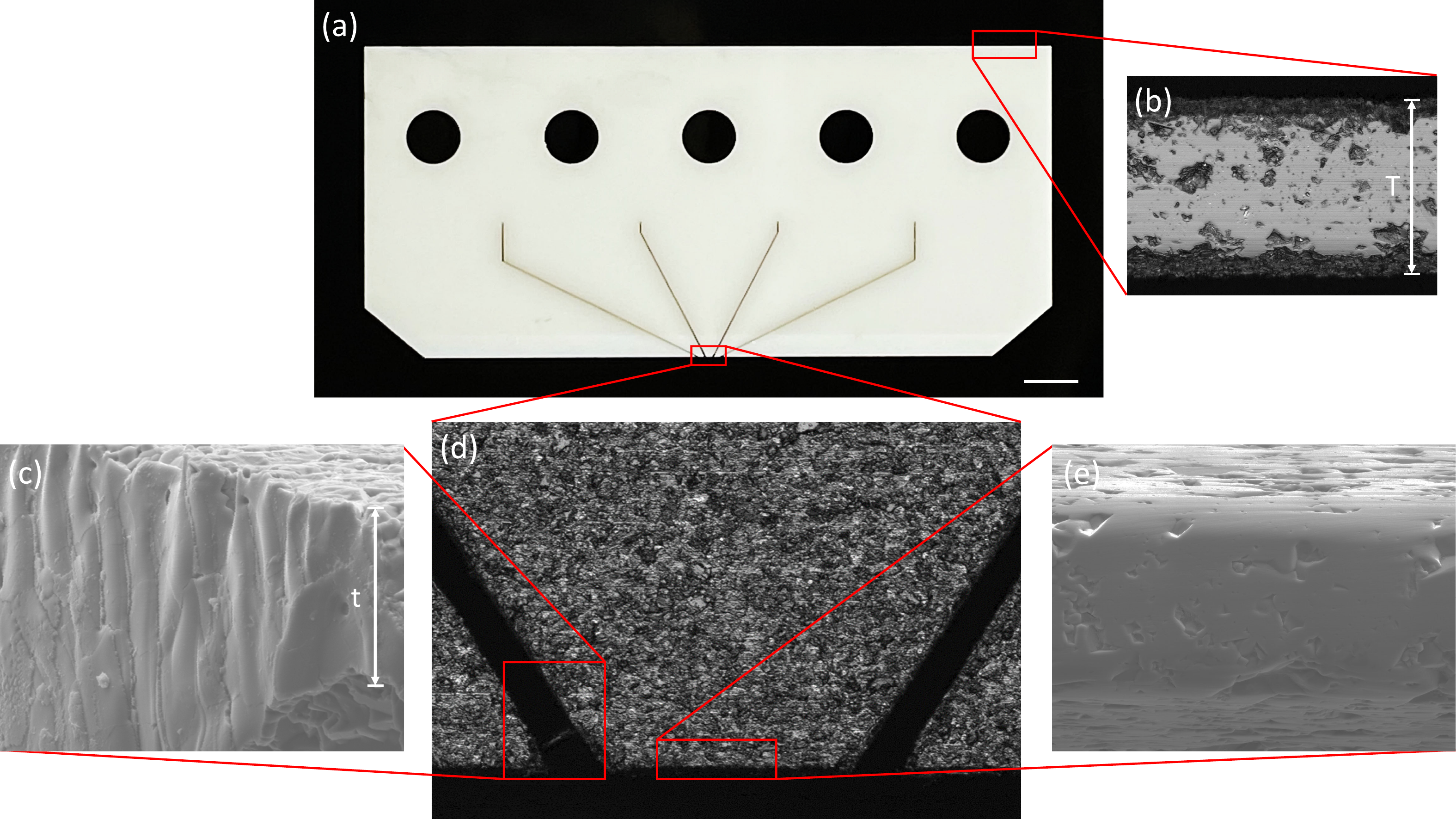} 
	\caption{
	(a) Photograph of machined alumina blade.
	Scale bar refers to 2~mm. 
	(b) Optical microscope image of non-tapered edge.
	$T = 300$~$\mu$m.
	(c), (e) Scanning electron microscope image of groove and tapered edge. 
	$t = 50$~$\mu$m.
	(d) Optical microscope image of central zone.	
	}
	\label{fig:blade_image}
\end{figure*}

\subsubsection{X-ray diffraction}

In order to characterize our alumina, we first examine the alumina substrate with an XRD spectroscopy.
The XRD is an extensively used technique to assess the physical properties of a solid, including the crystallinity, strain, defects, and crystal orientations. 
In formation of a bulk alumina from powders, the alumina exists in a series of different polymorphs: amorphous alumina is formed when the thermal decomposition occurs below $450\,^{\circ}\mathrm{C}$, seven metastable phases ($\gamma, \delta, \kappa, \rho, \eta, \theta$ and $\chi$) are generated at temperatures from 450 to 1,200$\,^{\circ}\mathrm{C}$, and thermally stable $\alpha$-alumina is created by sintering above 1,200$\,^{\circ}\mathrm{C}$~\cite{Wefers1987, Liu2005}.
Our XRD measurement (Fig.~\ref{fig:xrd_loss_tangent}(a)) shows that the diffraction peaks are clearly resolved, indicating our substrate is well crystallized. 
In this measurement, the uncertainty of the angle is $2\cdot 10^{-4}$ degrees and that of the intensity is dominated by the square root of the diffracted X-ray intensity, which is negligible.
Also, from the angles of the diffraction peaks, we identify that it is an $\alpha$-alumina~\cite{Wefers1987} with the crystal indices denoted in Fig.~\ref{fig:xrd_loss_tangent}(a).
To be used as a substrate, the $\alpha$-alumina is an appropriate choice, since the surface of $\alpha$-alumina is relatively less porous~\cite{Barma2014}, the bandgap is as large as 8.8~eV~\cite{French1990, Filatova2015}, and the dielectric loss is low (see below). 

\begin{figure*} [!t]
    \centering
	\includegraphics[width=5.5in]{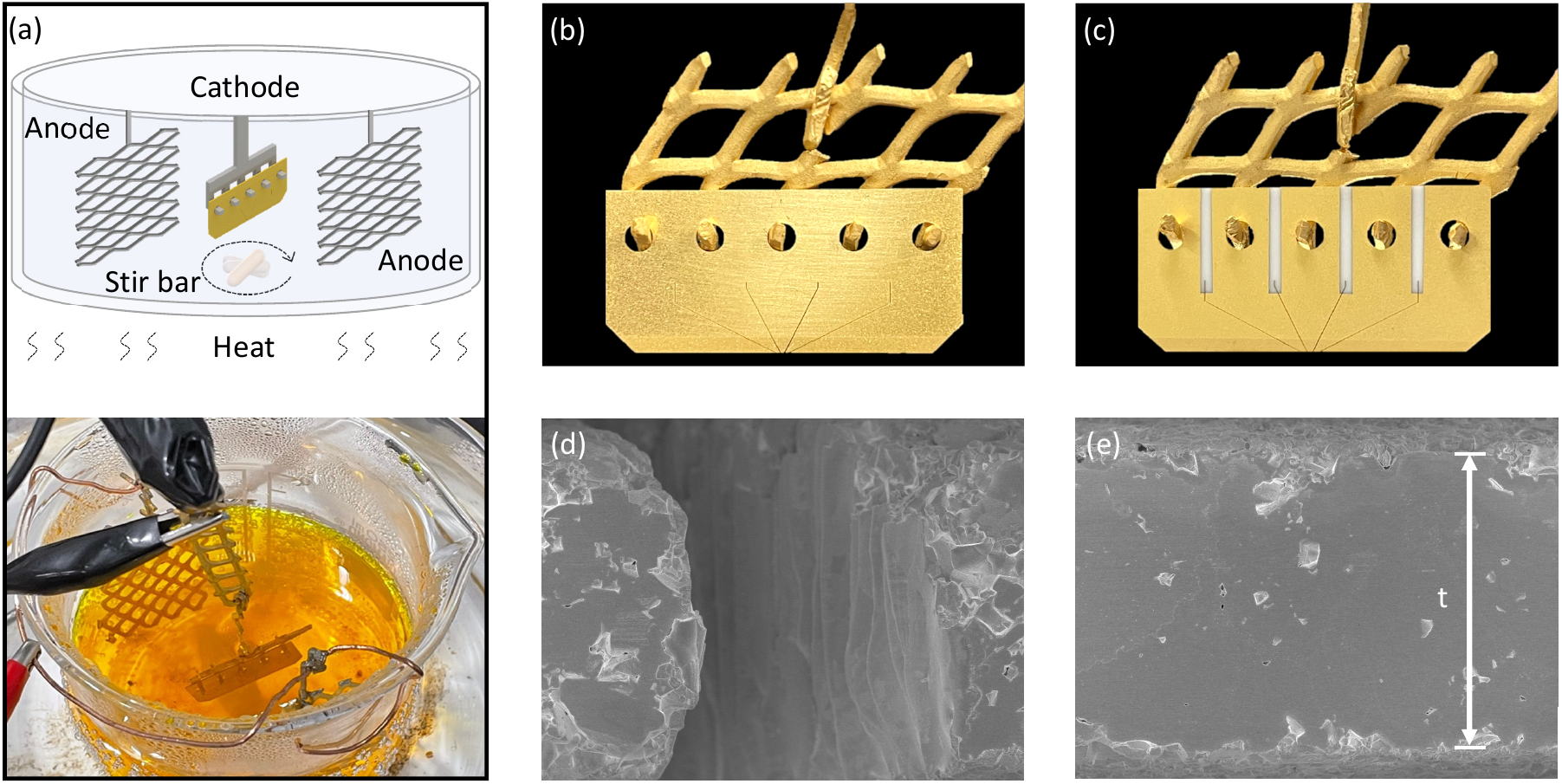} 
	\caption{
	(a) Schematic and photograph of electroplating setup. 
	Photographs of gold-plated RF blade in (b) and DC blade in (c) held with ``anchor''.
        Scanning electron microscope image after gold plating, on sidewall at groove in (d) and on tapered edge in (e). 
        $t=50$~$\mu$m.
	}
	\label{fig:gold_coating}
\end{figure*}

\subsubsection{Loss tangent}

Next characterization consists of a loss-tangent measurement. 
The loss tangent is a dimensionless value that quantifies the loss effect of an electromagnetic wave in a dielectric medium. 
Given a complex permittivity $\varepsilon$ of a dielectric, it follows that $\varepsilon =  \epsilon (1 + i \tan{\delta})$ with a real-number permittivity $\epsilon$ and loss tangent $\tan{\delta}$~\cite{Kumph2016}.
We perform the loss-tangent measurement using a precision impedance analyzer (Agilent Inc., 4294A). 
This device measures both the resistance and reactance at a given frequency, which returns the impedance and phase angle (loss tangent) consequently. 
Fig.~\ref{fig:xrd_loss_tangent}(b) shows the measurement results for four different alumina substrates. 
In our substrates, the overall tendency is that the loss tangent varies between $1.5 \cdot 10^{-2}$ to $2.5 \cdot 10^{-2}$ as the driving frequency increases from 10~kHz to few MHz, then the measured value decreases below $10^{-2}$ as the frequency increases to tens of MHz. 
The measurement uncertainty is $8.4\cdot10^{-3}$ of the obtained values, governed by the intrinsic voltage noise of the impedance analyzer; the error bar is  smaller than the symbol size of Fig.~\ref{fig:xrd_loss_tangent}(b).
In our trap driving frequency of 19.525~MHz, the loss tangent is measured to be about 0.01.
Our data are at a similar level with the results in Ref.~\cite{Vila1998}, however larger than the literature values in Refs.~\cite{Auerkari1996, Dolezal2015, Kumph2016}.
The difference would be attributed to several factors during the sintering, like temperature, pressure, and impurities in forming a bulk alumina from powders.
While we directly measure the loss tangent over a range of frequencies, indirect measurements can also give a relevant information: The dissipation factor can be inferred from scaling effects~\cite{Teller2021}, and the maximal value may be given by a trap's quality factor~\cite{Stick2006}.


\subsubsection{Machining}

We present the machining process of an alumina blade in Fig.~\ref{fig:fab_step}.
This engineering was performed in 21st Century, Co.~Ltd., Korea.
We make use of an alumina with a purity of 99.7\%.
The overall shaping shown in Figs.~\ref{fig:fab_step}(a)--(c) is done using a diamond grinder. 
Note that the tapering along one edge of the blade (rectangle in Fig.~\ref{fig:fab_step}(b) and Fig.~\ref{fig:fab_step}(c)) allows us to image or address the ion with a numerical aperture (NA) of 0.39 along one direction and 0.80 in the other direction, at a given blade-blade distance of 500~$\mu$m and the angle given in Ref.~\cite{Hong2022}.
After such shaping, the grooves and holes are made with laser micromachining (Figs.~\ref{fig:fab_step}(d) and (e)). 
The grooves are used for electrode segmentation, and the holes for mounting the blade with screws.
The deployed light source is a CW fiber laser with a power of 300~W operating at 1064~nm.
While the alumina body is exposed to the laser, the melted or evaporated ceramic residues are blown with nitrogen gas. 
The laser machining speed is about $0.4$~mm/min, resulting in a total elapsed time of 2~hours per a blade.
This speed is found empirically, which gives a best surface quality of the sidewall.
Fine polishing of the blade edge (only along the tapered side) follows as the last step of the blade machining.

Fig.~\ref{fig:blade_image} shows microscope images of a machined blade, which are observed after the cleaning process described in Sec.~\ref{sec:gold_coating}.
We point out three notable features. 
First, the overall nonuniform surface is a consequence of sintering, and this nonuniformity varies significantly by changing the sintering temperatures~\cite{Ahmad2010, Barma2014}.
Second, in Fig.~\ref{fig:blade_image}(c), we find the half-cylindrical shapes on the sidewall of the groove. 
It is because, while the alumina is melted by the high-power laser and hardened, the waist of the laser ($\sim15$~$\mu$m) is approximately ``mapped'' to this sidewall. 
Third, as shown in the bottom left (red rectangle) of  Fig.~\ref{fig:blade_image}(d), we sometimes observe the thread-like feature that exists between the grooves, probably due to the imperfect laser machining.
This would be problematic after plating with gold, because a short between the adjacent blade and spurious capacitance/inductance would be formed.
We therefore disconnect it by inserting a thin thread between the grooves; in the DC blade, this connection could also be broken by applying a high current between the shorted blades after gold coating.  
In Sec.~\ref{sec:surface_roughness}, we discuss the surface roughness of the blade in more detail.

\subsection{Gold coating}
\label{sec:gold_coating}

After the blade machining described above, we clean the alumina blade to remove ceramic residues and other contaminants. 
The cleaning is carried out in two stages, wet Piranha cleaning succeeded by dry oxygen plasma etching. 
We perform the Piranha cleaning for 30~minutes, where the blade is immersed in a solution of 4:1 mixture of sulfuric acid (H$_{2}$SO$_{4}$) and hydrogen peroxide (H$_{2}$O$_{2}$) at $100\,^{\circ}\mathrm{C}$. 
Next, we make mild oxygen-plasma etching on both sides of the blade, taking 1 minute for each facet.

Next, we continue with the metal sputtering, for an adhesion layer of a 30-nm thick titanium and a gold layer with a thickness of 200~nm. 
A dc sputtering is performed once for each blade side --- this causes, on the tapered edge and other three edges along the blade, four layers of titanium, gold, titanium, and gold are deposited sequentially. 
The thicknesses of those layers are 30/200/30/200~nm.
The deposition speed is 2.7~nm/min for titanium and 18~nm/min for gold. 
While the RF blade is sputtered without a mask, for the DC blade, a metal mask is used to block the deposition between adjacent segments (four white zones in Fig.~\ref{fig:gold_coating}(c)). 
After the sputtering, possible dirts are removed by a mild oxygen plasma cleaning, which would facilitate the next step of gold electroplating.

\begin{figure} [!b]
	\includegraphics[width=3.6in]{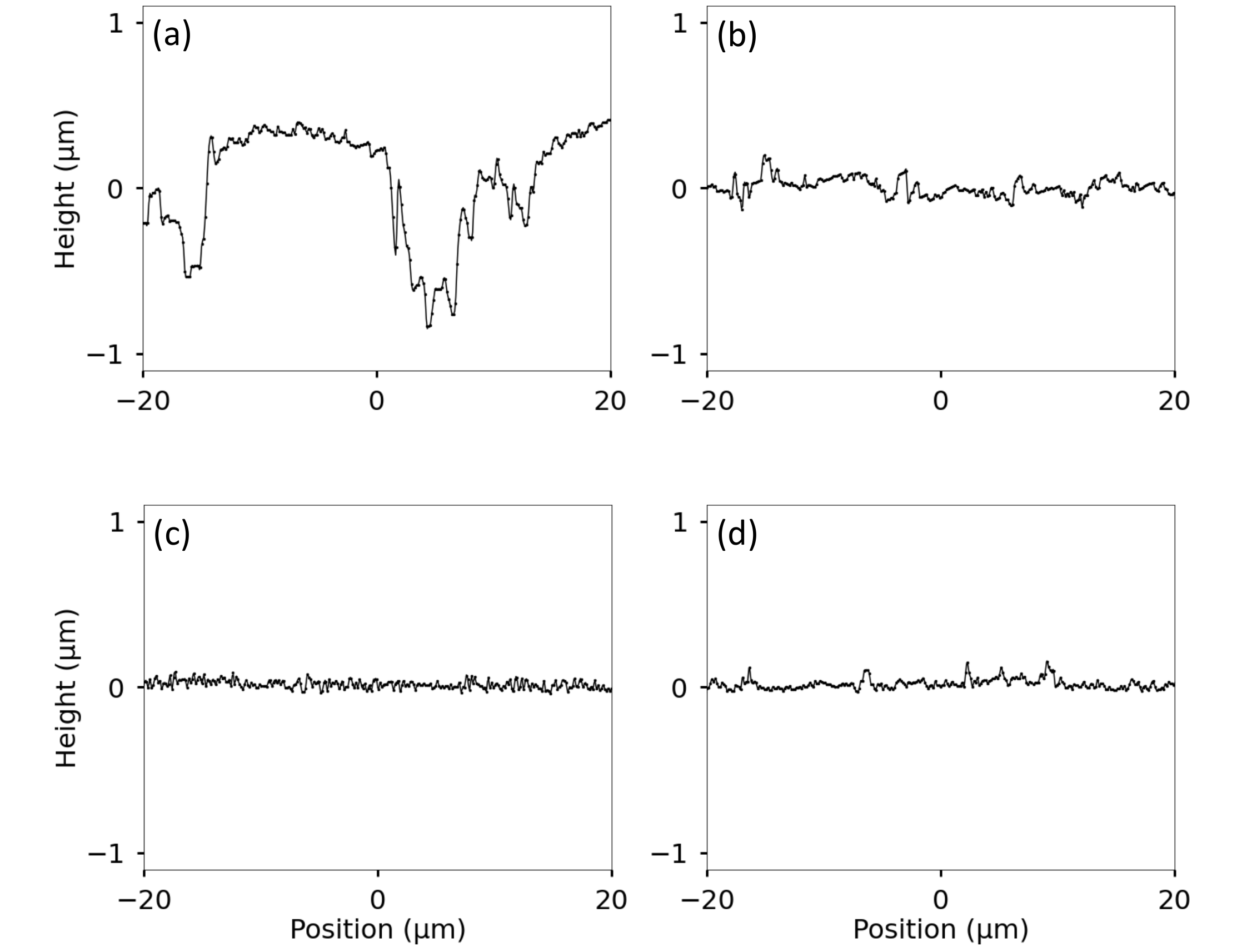} 
	\caption{
	Measured surface roughness on the tapered edge, after (a) laser machining, (b) polishing, (c) sputtering, and (d) gold plating.
	}
	\label{fig:roughness}
\end{figure}

\begin{figure*} [!t]
    \includegraphics[width=5.5in]{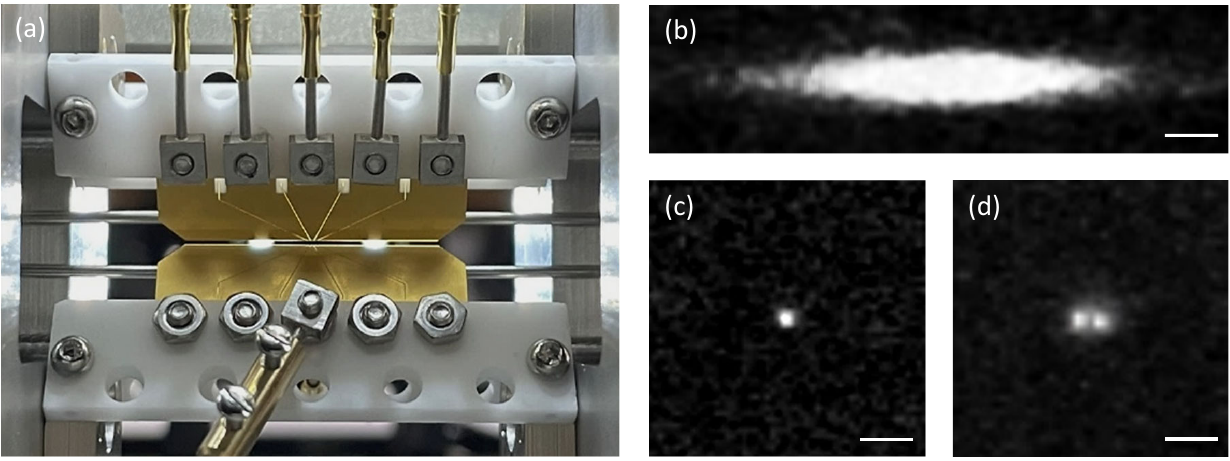} 
    \caption{
	(a) Photograph of our ion-trap system. 
	Fluorescence image of $^{174}$Yb$^{+}$ ion cloud in (b), single ion in (c), and two ions in (d).
	Scale bars denote 20~$\mu$m.   
    }
	\label{fig:trap}
\end{figure*}

Fig.~\ref{fig:gold_coating} presents our electroplating setup and SEM images of the completed RF and DC blades. 
In the setting shown in Fig.~\ref{fig:gold_coating}(a), we hold the blade with a customized, platinum-titanium ``anchor'' structure, that makes a simultaneous electrical contact of all five segments. 
The other end of the anchor is connected to the ground of a current-stabilized power supply, corresponding to the cathode of our plating system.  
The immersed parts of the anode are also made of platinum-titanium mesh, placed at both sides of the blade: this arrangement would assist a uniform plating over the whole surface. 
With a non-cyanide gold solution (JDC Inc., Temperesist K91S), we carry out a dc electroplating at a voltage 0.31~V and current 10~mA.
The plating is done at $65\,^{\circ}\mathrm{C}$ for 12 minutes while gently stirring the solution with a magnetic bar, resulting in a target thickness of 3~$\mu$m. 
After the plating, the blade is cleaned with deionized water, acetone, isopropyl alcohol, and dried with a nitrogen gun. 
The blade is stirred for 1 minute in each solution at $60\,^{\circ}\mathrm{C}$.
The gold plated surface is investigated using the SEM, as shown in Figs.~\ref{fig:gold_coating}(d) and (e). 
The whole surface is well covered with gold, and we provide quantitative descriptions in Sec.~\ref{sec:surface_roughness}.

\subsection{Surface roughness}
\label{sec:surface_roughness}

We discuss the surface roughness measurements.  
After each step of the blade fabrication, i.e. laser machining, polishing, sputtering, and electoplating, we make use of a three-dimensional confocal scanning microscope (Lasertec Inc., OPTELICS HYMRID C3) for the surface topography.
The surface roughness is quantified using the parameter $R_a = \sum_{i}^{N} |z_{i}-\bar{z}|/N$, where $N=283$ is the total measurement points, $z_{i}$ is the sample height at the measurement point $i$, and $\bar{z}$ is the average of all $z_{i}$.
The uncertainty of $z_{i}$ is about 1~nm dominated by mechanical stability of the microscope, and subsequently the error of $R_a$ is negligible.
The results are shown in Fig.~\ref{fig:roughness}.
Denoting the smallest $R_a$ values for the first three fabrication steps, we find 1.189~$\mu$m after laser machining (Fig.~\ref{fig:roughness}(a)), 41~nm on polished surface (Fig.~\ref{fig:roughness}(b)), and 20~nm after sputtering (Fig.~\ref{fig:roughness}(c)) respectively, over a scanning distance of 40~$\mu$m.
The decrease of the roughness from Fig.~\ref{fig:roughness}(b) to Fig.~\ref{fig:roughness}(c) would be attributed to the oxygen plasma cleaning.

The roughness on the gold-plated facet, directly seen by the ions, is investigated more carefully. 
We measure $R_a$ for randomly chosen several positions on the tapered edge, from which we obtain 24, 36, 50, 122, 136, and 188~nm over a scan length of 40~$\mu$m; the case of $R_a=24$~nm is presented in Fig.~\ref{fig:roughness}(d).
We remark the following points regarding this measurement. 
First, the roughness of electrodeposits would be dominantly determined by that of alumina substrate underneath.
Thus, in order to obtain great surface quality, it would be crucial to perform a high-quality sintering and to polish the alumina as finely as possible. 
Second, the high $R_a$ values, like $>100$~nm, on the gold-plated facet would be mostly originated from the small ``dents'' that appear in Fig.~\ref{fig:gold_coating}(e).
For removing these dents, the polishing should have been done more strongly, which takes the risk of breaking the corners of each segment. 
We also find that the surface quality is generally better on the central part, rather than both corners of the segment.
Lastly, we carry out an energy dispersive X-ray spectroscopy (EDS)~\cite{Goldstein2017} on the individual dents: gold is plated well at these locations.  
We define the ``purity of gold'' as a ratio between the integration of the peak associated with gold divided by that of all components. 
As a reference, EDS was performed on the several spots on a flat surface, an averaged purity is obtained as 98(1)\%; the mean of the purities at the dents is 95(5)\%, close to the reference value.

We compare the results of surface roughness with other previous works. 
In Ref.~\cite{Leibrandt2009}, the peak-to-peak roughness values of metal electrodes' surfaces, electrode edges, walls of a slot were reported as $\sim100$~nm, $300-500$~nm, and $1-5$~$\mu$m, respectively. 
The roughness of a laser-machined substrate was also investigated in Ref.~\cite{Ragg2019}: It is notable that after thermal polishing, microroughness of 1~nm and a flatness of $10-100$~nm was obtained. 
In Ref.~\cite{Hou2021}, a control of surface roughness was demonstrated between $R_{a}=6.2$~nm and 45~nm. 
Our work shows a similar surface quality with Refs.~\cite{Ragg2019, Hou2021}, and additionally provides step-by-step information over the course of trap fabrication.

\section{Ion-trap operation}

In Fig.~\ref{fig:trap}(a), we present our ion-trap apparatus made of the gold-plated alumina blades.  
Each blade is mounted to an alumina adapter that is fixed to the surrounding titanium holder. 
The blade-blade distance is 500~$\mu$m at an angle given in Ref.~\cite{Hong2022}.
The blades are aligned and assembled by hand, while monitoring the relative position with an optical microscope.
Two biasing rods, made of tungsten with a diameter of 1~mm, are shown behind the blades in Fig.~\ref{fig:trap}(a).

We point out that our chamber is ``passivated'' for reducing the outgassing rate.
The empty chamber (viewports and feedthroughs replaced with blank flanges) is pumped and baked at $450\,^{\circ}\mathrm{C}$ for 48~hours, with an oxygen partial pressure about 10$^{-9}$~torr.
Under such circumstance, an oxide film of Cr$_{2}$O$_{3}$ is formed all over the inner surface of stainless steel: this layer plays the role of a diffusion barrier for hydrogen, as well as provides a relatively uniform surface~\cite{Nuvolone1977, Ishikawa1996, Cho2000}. 
After the vacuum thermal oxidation, we install the ion-trap assembly in the chamber, bake the setup at $120\,^{\circ}\mathrm{C}$, and obtain a base pressure of mid $10^{-11}$~torr, probably due to the residual outgassing of in-vacuum components.
Since the base pressure decreases asymptotically, we might see the effect of the passivation from the generation rate of dark ions~\cite{Pagano2018, Obsil2019} or ion storage times, once the pressure would reach below $10^{-11}$~torr.

After placing the trap, the trap capacitance is measured to be $18$~pF using a LCR meter, through the connected electrical feedthrough. 
In conjunction with a helical RF resonator (quality factor about 270), we find a resonance frequency at 19.525~MHz, where the trap drive voltage is applied. 
The ions are imaged with an objective lens with a NA of 0.38, followed by a tube lens of a focal length 200~mm: the resultant magnification is 8.25 at 369~nm.
After the two-photon ionization of $^{174}$Yb$^+$ ions with 399~nm and 369~nm lasers, we load a laser-cooled $^{174}$Yb$^+$ cloud in Fig.~\ref{fig:trap}(b), and single and two ions in Figs.~\ref{fig:trap}(c) and (d). 
The laser frequencies of two lasers, and that of a repump laser at 935~nm are simultaneously stabilized using the scheme in Ref.~\cite{Kim2021}.
When the ion-fluorescence images in Figs.~\ref{fig:trap}(c) and (d) are taken, the trap frequencies are measured by exciting the ion motion with auxiliary driving frequencies: we obtain $3.2$~MHz along the radial directions, and 350~kHz along the trap axis. 

\section{discussion}

While the alumina was often used as a substrate for ion-trap devices~\cite{Hensinger2006, Blakestad2009, Kienzler15, Kaufmann2017, Lu2019}, it might not be an optimal choice from the perspective of laser machining.
It is because inefficient absorption requires more laser power, and random scattering causes more rough surface after the engineering.  
Instead, by adding small amount of inorganic pigments to a ceramic powder (before sintering), the color of the material becomes black, which enhances the light absorption without substantial change of the physical properties~\cite{Slocombe2000}.
For instance, black-colored alumina or zirconia, could be chosen for the ion-trap substrate. 
We note that the dielectric loss of zirconia is as small as that of alumina; however, zircona has rather low thermal conductivity ($5\sim7$~W/(m$\cdot$K)) that would induce more heating of the RF blade while operating the ion trap.  
Thus zirconia would be suitable for a more miniaturized trap that may require weaker RF power, or a trapping system for lighter ions like beryllium and calcium than ytterbium.

We envisage future experiments that connect the substrate properties with ion measurements.
Concerning the loss tangent, since there are many sources of the frequency shift of the qubit transition, it would be difficult to measure the BBR shift separately. 
However, if one would measure a change of the frequency shift as the RF driving power varies, one might extract information about the dielectric loss: The transition might shift more as the RF power increases.
For surface roughness, it has been known that the roughness affects the heating rate of the ions~\cite{Hite2012, Brownnutt2015}. 
As studied in Sec.~\ref{sec:surface_roughness}, the surface is more uniform at the center than the corner of the segment; if one measures a change of the ions' heating rate at several positions along the trap axis, we would expect that, e.g.  the heating rate might be higher when the ion is close to the corner of the segment.


\section{conclusion}

In summary, we have described a development process of a segmented-blade linear ion trap.
The XRD spectroscopy shows that the substrate is an $\alpha$-alumina, and the loss tangent is measured to be about $10^{-2}$ at $20$~MHz.
The alumina blade is micromachined with a laser, succeeded by polishing and gold-coating. 
We quantify the surface roughness at every step of the fabrication, and our confocal optical microscopy gives a height deviation as small as $<30$~nm on a gold-plated surface. 
In our ion-trap apparatus made of those blades, trapping of laser-cooled $^{174}$Yb$^{+}$ ions is demonstrated successfully.
We expect that our detailed characterization of the substrate would assist the development of a state-of-the-art ion-trap hardware, in particular for dealing with the BBR shift of the qubit transition and surface-induced heating effects.  

\begin{acknowledgments}
We thank Jae Kap Jung at Korea Research Institute of Standards and Science for the help in the loss-tangent measurement. 
This work has been supported by  BK21 FOUR program and Educational Institute for Intelligent Information Integration, National Research Foundation (Grant No.~2019R1A5A1027055), Institute for Information \& communication Technology Planning \& evaluation (IITP, Grant No.~2022-0-01040), Samsung Science and Technology Foundation (SSTF-BA2101-07) and Samsung Electronics Co., Ltd. (IO201211-08121-01).
\end{acknowledgments}

\section*{Data Availability Statement}
All the data in this work are available in https://zenodo.org/record/7096020\#.YzeN3XZByUk~\cite{Kim2022}.

\bibliographystyle{apsrev4-2}
\bibliography{bibliography}

\end{document}